\documentclass[9pt,twocolumn,twoside]{optica}

\usepackage{tablefootnote}

\setboolean{shortarticle}{true}
\setboolean{minireview}{false}

\title{Overcoming unambiguous state discrimination attack with the help of Schr\"odinger Cat decoy states}

\author[1]{Andrei Gaidash}
\author[1,*]{Anton Kozubov}
\author[1]{George Miroshnichenko}

\affil[1]{Department of Photonics and Optical Information Technology, ITMO University, Saint Petersburg, Russia}

\affil[*]{Corresponding author: avkozubov@corp.ifmo.ru}

\begin{abstract}
In this work we propose the technique for phase-coded weak coherent states protocols utilizing two signal states and one decoy state which is found as linear combination of signal states (Schr\"odinger Cat states); the latter allows to overcome the USD attack. For instance, Schr\"odinger Cat states can be considered as even coherent states. Moreover we consider decoy states implementation based on squeezed vacuum states which might not disables USD completely yet produces discrimination probabilities low enough to distribute keys in channel with particular losses. Thus we can detect Eve simply by monitoring the detection rate of decoy states. It should be noted that this approach can be scaled to more complex schemes.
\end{abstract}

\setboolean{displaycopyright}{true}

\begin{document}

\maketitle

\section{Introduction}

Quantum key distribution is extremely important field of secure communication and its security is provided by the quantum laws. For the last decades huge amount of different theoretical and experimental works have been performed where a lot of various schemes with weak coherent sources instead of true single photon sources were suggested due to their utility, for instance \cite{weak2,weak3}. Some of them use phase coding protocols instead of the ones based on the polarization, for instance \cite{weak1ph,weak2ph,Miroshnichenko18,Gleim16,Gleim17}. In fact each phase coded quantum states, although they can be generated not in a usual way \cite{Miroshnichenko17,Miroshnichenko18}, in the set have overlapping with each other and that gives a crucial opportunity to Eve to provide unambiguous state discrimination (USD) \cite{Lutkenhaus00,Chefles98,Chefles04,Barnett98,Ivanovic87,Peres98}.
 
As far as weak coherent states are overlapped, they can not be distinguished perfectly, so Eve can discriminate them unambiguously only with the arbitrary probability (depending on the values of overlapping). In this work we consider only USD attack with inconclusive result yet zero error. This means that Eve has only two possible outcomes for each state: the unambiguous identification of the particular state and the inconclusive result. All inconclusive results can be blocked by Eve and she can increase the intensity of sent pulses to maintain the detection rates. Also she can perform additional bitflips in order to maintain error rates since amplified signal will produce less errors. The strategy allows Eve to provide zero-error attack maintaining both raw key rates and error rates.

Here we would like to consider the protocol which utilizes only two weak coherent signal states.  Alice sends two states with phase $\varphi_A\in\{0,\pi\}$; the first state encodes the logical 0 while the second encodes the logical 1. The set of phases used by Bob is the same as that of Alice. Decoding is based on the fact that each time Alice and Bob randomly chose the same phase there should be detection event and in opposite case there should not be one (except for the dark counts). In this work we propose the special decoy state which Alice sends with much more lower a priori probability than the signal states. Decoy states only provide possibility to detect the USD attack and are not used to extract the secret key out of them.

In this paper we propose the solution to overcome USD attack on phase protocols with weak coherent pulses utilizing particular kind of decoy states - Schr\"odinger cat states. This paper organized as follows. Section 2 gives a detailed general description of Eve's strategy. In section 3 we describe unambiguous state discrimination measurement and show fundamental impossibility to discriminate the states for Eve. Section 4 dedicated to symmetrical case which is natural for QKD systems and proposed Schrödinger cat states as decoy states. Particular case is considered in Section 5 where we analyze squeezed vacuum decoy states. In Section 6 we provide our results and conclude the article.

\section{The attack}
\begin{figure}
\begin{center}
\includegraphics[width=\linewidth]{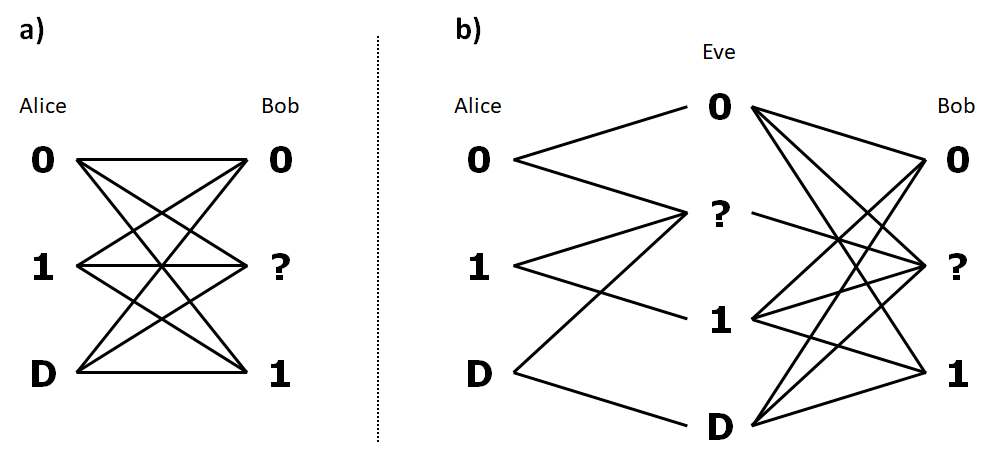}
\end{center}
\caption{a) The Alice-Bob channel . b) The Alice-Eve-Bob channel, where Eve performs USD attack. It may be considered as two consequent Alice-Eve and Eve-Bob channels. Symbols "$0$", "$1$", "$D$" denote states responsible for logical $0$ and $1$ together with decoy states respectively. Symbol "$?$" denotes the inconclusive result. Conditional probabilities between input and output parameters of the Alice-Bob and Alice-Eve-Bob channels can be found in Tables~\ref{tab:AB channel} and~\ref{tab:AEB channel} respectively. Combined channel's relations can be found as linear combination of the Alice-Bob and Alice-Eve-Bob channels' parameters with probability $1-P_e$ and $P_e$ respectively.}
\label{fig:false-color}
\end{figure}

We would like to consider the USD attack with respect to the scheme shown in Fig.~\ref{fig:false-color}. Assuming Eve can split channel in two - the Alice-Bob channel and the Alice-Eve-Bob channel (lossless one or with lower losses) where she performs the USD measurements. She decides how many states goes in each channel; let us denote the probability of state going to Eve's channel to be $P_e$.

The original Alice-Bob channel is the classical-quantum Markov cryptographic symmetric binary error and erasure channel (BSEE) with probabilities of the inconclusive result $G$ and the bitflip $E$ (and the conclusive result denoted as $C=1-G-E$) respectively for signal states. However we also include arbitrary decoy state which has some probabilities $D^{(0)}$ and $D^{(1)}$ to be interpreted as logical 0 and 1 respectively. For simplicity one may denote the probability of decoy state detection as $D=D^{(0)}+D^{(1)}$ however in particular case one may observe them separately. Conditional probabilities of all possible outcomes are denoted in Table~\ref{tab:AB channel}.

The altered Alice-Eve-Bob channel can be considered as combination of the binary symmetric erasure channel  (BSE) at the Alice-Eve part and the BSEE at the Eve-Bob part with additional decoy state input. The first part is characterized with probabilities $P_S$ and $P_D$ related to USD for signal and decoy states respectively. The second part is the same as the original channel except for the modified parameters (denoted with the lower indices $e$) and the input of inconclusive results from the Alice-Eve part. If Eve performs USD only on the part of qubits then combination of the Alice-Bob (with probability $1-P_e$) and Alice-Eve-Bob (with probability $P_e$) channels' parameters are denoted by tildas. Calculated relations between input and output parameters (i.e. conditional probabilities) for this channel can be found in Table~\ref{tab:AEB channel}.

We would like to consider the case where Eve tries to keep the detection rates and the error rates the same as in the original channel (like without USD attack) as follows
\begin{eqnarray}
\tilde{D}=D+P_e\left(P_DD_e-D\right)=D, \\
\tilde{G}=G+P_e\left(P_S(1-G_e)-(1-G)\right)=G, \\
\tilde{E}=E+P_e\left(P_SE_e-E\right)=E.
\end{eqnarray}
Let us consider further only decoy states since it is free parameter that will not affect the raw key generation rates ($1-G$). As one may notice the only way for legitimate users to prevent USD attack is to choose decoy state that even if $D_e$ approaches to unity the inequality condition as follows is satisfied
\begin{equation}\label{PDD}
P_D<D.
\end{equation}
This relation helps us estimate maximal allowed losses in the channel since $D$ is dependent on them. It should be noticed that Eve's method of discriminating signal states is always better then Bob's one due to her ideal equipment. Nevertheless Eve have to discriminate also the decoy states in our case, and for Bob it is not necessary since Alice always tells him during the reconciliation which exact pulse or time bin contains the decoy. In this way, if the structure of decoy state will be correct we can provide fundamental impossibility to discriminate such states to Eve. So let us consider further USD measurement in more details and find suitable parameters on decoy states that satisfies Eq.~\ref{PDD}.

\begin{table}
\caption[\bf Conditional probabilities between input (vertical) and output (horizontal) parameters in the  Alice-Bob channel]{\bf Conditional probabilities between input (vertical) and output (horizontal) parameters in the Alice-Bob channel\footnote[1]{}.}
\begin{center}
\begin{tabular}{cccc}
\hline
Input/Output & $0$ & $1$ & Inconclusive result \\
\hline
$0$ & $C$ & $E$&$G$ \\
$1$ & $E$ & $C$& $G$\\
$D$ & $D^{(0)}$ & $D^{(1)}$& $1-D^{(0)}-D^{(1)}$ \\
\hline
\end{tabular}
\end{center}
  \label{tab:AB channel}
  \footnote[1]{}\label{1}$G$, $E$ and $C=1-G-E$ are probabilities of the inconclusive result,  the bitflip, and the conclusive result for signal states respectively, $D^{(0)}$ and $D^{(1)}$ are probabilities to interpret decoy state as logical $0$ and $1$ respectively.
\end{table}

\begin{table*}[htbp]
\centering
\caption[\bf Conditional probabilities between input (vertical) and output (horizontal) parameters in the Alice-Eve-Bob channel.]{\bf Conditional probabilities between input (vertical) and output (horizontal) parameters in the Alice-Eve-Bob channel\footnote[3]{}.}

\begin{center}
\begin{tabular}{cccc}
\hline
Input/Output& $0$ & $1$ & Inconclusive result  \\
\hline
$0$ & $(1-P_e) C+ P_e P_SC_e $ & $(1-P_e)E+ P_eP_S E_e$&$(1-P_e)G +P_e \Big( P_S G_e+(1-P_S) \Big)$ \\
$1$ & $(1-P_e)E +P_eP_S E_e$ & $(1-P_e)C+ P_eP_SC_e$& $(1-P_e)G +P_e \Big( P_S G_e+(1-P_S) \Big)$\\
$D$ & $(1-P_e)D^{(0)}+ P_eP_D D_e^{(0)}$ & $(1-P_e)D^{(1)} +P_eP_D D_e^{(1)}$& $(1-P_e) \Big( 1-D \Big) +P_e \Big( P_D(1-D_e)+(1-P_D) \Big)$ \\
\hline
\end{tabular}
\end{center}

  \label{tab:AEB channel}
  \footnote[3]{} $P_S$ and $P_D$ are probabilities to unambiguously discriminate signal and decoy states respectively, $G_e$, $E_e$, and $C_e=1-G_e-E_e$ are altered probabilities of the inconclusive result, the bitflip, and the conclusive for signal states respectively, $D^{(0)}_e$ and $D^{(1)}_e$ are altered probabilities to interpret decoy state as logical $0$ and $1$ respectively, $D_e=D^{(0)}_e+D^{(1)}_e$.
\end{table*}

\section{Unambiguous state discrimination}
Unambiguous state discrimination  was proposed firstly in \cite{Chefles98,Barnett98,Ivanovic87} and developed in the next papers \cite{Lutkenhaus00,Chefles04}. The main idea of this measurement is to unambiguously identify non-orthogonal linearly independent quantum states. First method of distinguishing three states was proposed in \cite{Peres98}. Let us consider three quantum states $|u_1\rangle$, $|u_2\rangle$ and $|u_3\rangle$. Two of them ($|u_1\rangle=|\alpha e^{i \phi}\rangle$ and $|u_2\rangle=|-\alpha e^{i \phi}\rangle$) supposed to be the signal states and can be taken as the weak coherent states with the amplitude $\alpha$ and the phase $\phi$. The third one ($|u_3\rangle$) is the decoy state and should be chosen correctly. Signal states should be sent in the channel with equal a priori probability $\frac{1-\nu}{2}$ and decoy state with a priori probability $\nu$. Those states could be unambiguously discriminated by Eve, if she correctly determines the positive-operator valued measure (POVM) as $\sum\limits_{i=0}^3\hat{A}_i=\hat{I}$ (where $\hat{I}$ is unity matrix). Here the operator $\hat{A}_0$ is related to an inconclusive result, which is always presented due to nonorthogonality of the states $|u_1\rangle$, $|u_2\rangle$ and $|u_3\rangle$. Operators $\hat{A}_i$ contain probabilities of successful discrimination of signal and decoy states respectively which are variable parameters. The detailed construction method of operators $\hat{A}_i$ was described in \cite{Peres98}. Let us denote the elements of overlapping matrix of our states
\begin{equation}\label{SU}
S_{12}=\langle u_1|u_2\rangle, S_{13}= \langle u_1|u_3\rangle, S_{23}=\langle u_2|u_3\rangle.
\end{equation}
The next step is to denote the orthonormal basis in three-dimensional vector space $|u_1\rangle$, $|u_2\rangle$ and $|u_3\rangle$ using the Gram-Schmidt orthogonalization process. In this orthonormal basis the vectors ($|u_i\rangle$) and the normalized reciprocal \footnote[2]{By reciprocal we mean relations $\langle v_i|u_j\rangle=0$ if $i\not=j$ and $\langle v_i|u_j\rangle=1$ if $i=j$.}  to them ($|v_i\rangle$) (also known as binormalized) are denoted as
\begin{eqnarray}
|u_1\rangle=
\begin{pmatrix}
    1 \\
    0 \\
    0
\end{pmatrix},
|u_2\rangle=
\begin{pmatrix}
    S_{12} \\
    L \\
    0
\end{pmatrix},
|u_3\rangle=
\begin{pmatrix}
    S_{13} \\
   \frac{K}{L}  \\
   \frac{M}{L} 
\end{pmatrix}, \\
|v_1\rangle=
\begin{pmatrix}
    1\\
    \frac{-S_{12}^*} {L} \\
   \frac{H^*}{LM}
\end{pmatrix},
|v_2\rangle=
\begin{pmatrix}
    0 \\
   \frac{1}{L}\\
    \frac{-K^*}{LM}
\end{pmatrix},
|v_3\rangle=
\begin{pmatrix}
   0\\
   0\\
    \frac{L}{M}
\end{pmatrix},
\end{eqnarray}
where $H=S_{12}\cdot S_{23}-S_{13}$, $K=S_{23}-S_{12}^* \cdot S_{13}$, $L=\sqrt{1-|S_{12}|^2}$, $M=\sqrt{1-|S_{13}|^2-|S_{12}|^2-|S_{23}|^2+ S_{12}^* S_{13} S_{23}^*+S_{12} S_{13}^*S_{23}}$.
Here one may notice that if $|u_3\rangle$ can be expressed as normalized linear combination of $|u_1\rangle$ and $|u_2\rangle$ then $M=0$ and the construction of $|v_i\rangle$ vectors is impossible. Thus USD measurement cannot be implemented by Eve and we can overcome the USD attack.
\section{Symmetrical case}
Let us consider symmetrical case which is naturally takes place in QKD systems (however following can be extended to more general case if necessary) where
\begin{equation}\label{SU}
\begin{split}
S_{12}&=\langle u_1|u_2\rangle=\exp(-2\alpha^2),  \\ 
S_{13}&=S_{23}=\langle u_1|u_3\rangle=\langle u_2|u_3\rangle,
\end{split}
\end{equation}
where $\alpha$ is amplitude of the coherent state. Operator $\hat{A}_0$ is denoted as
\begin{equation}\label{A0}
\hat{A}_0=\hat{I}-\Big( P_S \cdot (|v_1\rangle \langle v_1|+|v_2\rangle \langle v_2|)+P_D \cdot |v_3\rangle \langle v_3| \Big),
\end{equation}
where $P_S$ is equal probability of successful discrimination for both signal states and $P_D$ is probability of successful discrimination for decoy states.
In the orthonormal basis the operator matrix $\hat{A}_0=\hat{I}-\sum\limits_{i=1}^3\hat{A}_i$  should be represented as follows
\begin{eqnarray}
\hat{A}_0=
\begin{pmatrix}
1-P_S & \frac{P_S \cdot S_{12}}{L} & \frac{-P_S \cdot H }{L \cdot M}\\
\frac{P_S \cdot S_{12}^*}{L} & 1-\frac{P_S\cdot (1+|S_{12}|^2)}{L^2} & \frac{P_S\cdot ( S_{12}^* \cdot H + K)}{M \cdot L^2} \\
\frac{-P_S\cdot H^*}{L \cdot M}  & \frac{P_S \cdot ( S_{12} \cdot H^* + K^*)}{M \cdot L^2} & 1-\frac{P_S\cdot \left( |H|^2 + |K|^2 \right)+L^4\cdot P_D}{(M \cdot L)^2}
\end{pmatrix}.
\end{eqnarray}
The determinant of $A_0$ is derived as
\begin{eqnarray}
\begin{split}
\det(A_0)=\frac{1}{M^2}\cdot \Big( 2\cdot P_D \cdot P_S + P_S^2 - P_D \cdot L^2 - \\
- P_S\cdot(2-|S_{13}|^2-|S_{23}|^2) - P_D \cdot P_S^2 + M^2 \Big).
\end{split}
\end{eqnarray}

Let us denote the mean probability of inconclusive result as $P_0$. This probability depends on variable parameters and equals to 
\begin{equation}\label{P0}
P_0=1-(1-\nu)P_S-\nu P_D.
\end{equation}
The main goal of USD consists in solving the optimization problem to minimize the probability $P_0$. The minimum should be investigated when operator $A_0$ has zero eigenvalue. This condition corresponds to $\det(A_0)=0$ \cite{Peres98}. Then optimization problem has the next form
\begin{eqnarray}
\begin{cases}
P_0=1-(1-\nu)P_S-\nu P_D\\
\det(A_0)=0
\end{cases}.
\end{eqnarray}
Let us express $P_D$ from the second condition, $\det(A_0)=0$, as follows
\begin{equation}\label{PD}
P_D=f_1(P_S)=\frac{P_S-\Delta}{P_S-1-S_{12}},
\end{equation}
where $\Delta=1+S_{12}-2|S_{13}|^2$.
Also let us define $P_D$ from Eq.~\ref{P0} as follows
\begin{equation}
P_D=f_2(P_S)=\frac{1-P_0-(1-\nu)P_S}{\nu},
\end{equation}
when the value of $P_0$ is fixed.
Consider two curves $f_1$ and $f_2$ on the surface ($P_D, P_S$). Varying $P_0$ determines the intersection of these two curves. Obtained values of $P_S$ and $P_D$ are the optimal probability of the unambiguous discrimination of the signal and the decoy states respectively for given $P_0$ (nevertheless not all values of $P_0$ corresponds to intersection). 

In fact $P_S$ and $P_D$ can be equal to zero if we chose $|u_3\rangle$ in some special way. In this case decoy state should be described as the Schr\"odinger Cat state \cite{Kennedy88,Schleich91,Molmer06,Kim93} and looks like the linear combination of two weak coherent states
\begin{equation}
|u_3\rangle=\frac{|\alpha e^{i \phi}\rangle+|-\alpha e^{i \phi}\rangle}{\sqrt{2\left(1+\exp(-2\alpha^2 \right)}}.
\end{equation}
Utilizing Schr\"odinger Cat states (even coherent states) one can make USD attack useless at all.

\section{Squeezed vacuum realization}

Proposed decoy method may be implemented utilizing parametric generation; one may chose $|u_3\rangle$ as the single mode squeezed vacuum state \cite{Weedbrook12} as follows
\begin{eqnarray}
|u_3\rangle=|0,r\rangle=\frac{1}{\sqrt{\cosh(r)}}\sum_{n=0}^{\infty}\frac{\sqrt{(2n)!}}{2^n n!}\left(\tanh(r)\right)^n |2n\rangle,
\end{eqnarray}
where $r$ is the squeezing parameter. Further we consider real $r$ and $\varphi=0$. Thus according to Eq.~\ref{PD} following expression should be minimized varying the squeezing parameter $r$ in order to minimize $P_S$ and $P_D$
\begin{eqnarray}\label{delta}
\begin{split}
\Delta &= 1+\exp(-2\alpha^2)-\frac{2}{\cosh(r)}\exp\Big(-\alpha^2\big(1-\tanh(r)\big)\Big).
\end{split}
\end{eqnarray}
Minimum value of $\Delta$ is obtained when $\alpha$ and $r$ are related as
\begin{eqnarray}
\alpha=\sqrt{\exp(-r)\cdot \cosh(r) \cdot \ln\Big(\exp(r) \cdot \left(\cosh(r)\right)^2\Big)}.
\end{eqnarray}

Nevertheless squeezed vacuum decoy state might not disables USD completely yet produces discrimination probabilities low enough to distribute keys in channel with particular losses. In this case one should pay attention to the threshold on the number of detected decoy states $N_D$. If following inequality satisfies that there might be USD attack with the probability erf$(\frac{z}{\sqrt{2}})$, where the latter is error function
\begin{eqnarray}
N_D\le N\tilde D + z\sqrt{N\tilde D(1-\tilde D)}<N D-z\sqrt{N D (1-D)},
\end{eqnarray}
and where $N$ is the number of sent decoy states. 
This threshold provide useful rule of thumb estimating when QKD can be performed in the channel with maximal allowed losses $L_{max}$ equals approximately to difference between the maximal possible detection probability of decoy states and the probability of their discrimination in dB as follows:
\begin{equation}
L_{max}\approx -10\log_{10}(\mu \eta_B \eta_D - P_D),
\end{equation}
where $\mu$ is mean photon number for the decoy states, $\eta_B$ is losses on the Bob's side, and $\eta_D$ is quantum efficiency of the detector. 

\section{Conclusion}
In this work we consider the approach for QKD protocols with phase coded weak coherent states that disables USD attack utilizing Schr\"odinger Cat decoy states. We demonstrate the method considering only two signal states and one decoy state which is the simplest version of the protocol. However it is scalable and one may follow the article in order to apply the method for arbitrary amount of used states.

Symmetrical case following Eqs.~\ref{SU} and~\ref{A0} is naturally takes place in QKD systems so it is essential to consider it. Eq.~\ref{PD} allows us to chose decoy state $|u_3\rangle$ in order to nullify or minimize both $P_S$ and $P_D$. The Schr\"odinger Cat states (even coherent states) and squeezed vacuum states are responsible for the first and the second cases respectively.

\section*{Funding Information}
Ministry of Education and Science of Russian Federation (contract 03.G25.31.0229).

\section*{Acknowledgments}

All authors contributed equally to the work.


\end{document}